\documentstyle[12pt]{article}
\newcommand{\LI}{\hbox to\hsize}
\newcommand{\LLI}[1]{\LI{#1\hss}} \newcommand{\RLI}[1]{\LI{\hss#1}}
\newcommand{\CLI}[1]{\LI{\hss#1\hss}}

\newcommand{\PM}[1]%
{\mbox{$m_{\rm #1}$}} 
\newcommand{\BEQ}{\begin{equation}}
\newcommand{\EEQ}{\end{equation}}

\newcommand{\gappr}{\mbox{$\stackrel{>}{\sim}$}} 
\newcommand{\mr}[1]{\mbox{\rm #1}} 
\newcommand{\ack}{\LLI{\Large{\bf Acknowledgement}}
\vspace*{2mm}
\par
\noindent
\nopagebreak
This research has been supported in part by the U.S. Department of Energy
under Grant \#~DE--FG--02--85ER40211} 
\newcommand{\incircle}[1]{\mbox{{\hbox{$\bigcirc$}\kern-0.7em
\lower0.05ex\hbox{\mbox{{\scriptsize\rm #1}}}}}}
\newcommand{\eq}[1]{eq.~(\ref{#1})}
\newcommand{\ETC}{\mbox{\em etc.\/ }}

\newcommand{\CF}{\mbox{\em cf.\/ }}
\newcommand{\IE}{\mbox{\em i.e. \/}}
\newcommand{\ETAL}{\mbox{\em et. al.\/ }}
\newcommand{\EG}{\mbox{\em e.g.\/ }}

\begin{document}
\begin{titlepage}
\RLI{JHU--TIPAC--920027/rev}  
\RLI{May, 1993} 

\vspace*{2.0cm}
\begin{center}
{\Large\bf Astrophysical Limit on the Deformation
of the Poincar\'{e} Group}\\[2.0cm] 
G. Domokos and S.
Kovesi--Domokos\footnote{E-mail: SKD@JHUP.PHA.JHU.EDU}\\[5mm]
The Henry A. Rowland Department of Physics and Astronomy\\
The Johns Hopkins University\\
Baltimore, MD 21218 \\[2.5cm]
\end{center}
%
The  deformed Poincar\'{e} group
contains a characteristic mass, $\kappa$. At energies exceeding $\kappa$,
deviations from the special theory of relativity become significant.
However, small deviations from ordinary relativistic kinematics
are observable even at energies substantially lower than $\kappa$.
The observation of distant events producing ultra high energy (UHE) particles
leads to a lower limit on $\kappa$. From an analysis of the  UHE data
on the burst in the binary system HER X1 in 1986, we deduce
$\kappa \gappr  10^{12}$~GeV.
%
\vfill
\end{titlepage}
\newpage
The special theory of relativity is one of the fundamental ingredients of
any calculation or experiment in particle physics. Therefore, it is
important to establish the accuracy to which it can be trusted.

This is a  non--trivial problem. Practically all
analyses  of current particle physics experiments  assume the validity
of the special theory of relativity, \IE   the invariance of {\em local}
physics under the Poincar\'{e} group: hence, by design,
they cannot be used to test
Poincar\'{e} invariance, unless, perhaps, that invariance fails
in a dramatic and
truly unexpected way. As a consequence, while one knows that the
{\em special} theory of relativity is not applicable on length scales
where gravitational effects are important, its validity at short
distances is harder to test.

It has been proposed several times that Poincar\'{e}
invariance is violated at short distances; for some recent proposals
see~\cite{SN1987A}. However, most of the  proposals
were introduced in an {\em ad hoc} fashion; the internal
consistency of the  schemes has never been
investigated systematically.
(There also exists a substantial number of older works in this area,
involving a variety of ideas, such as the discretization of
space--time, \ETC; it is not our purpose  to review
them here.)

Due to  developments in the theory of group deformations,
the situation has changed, however. In some recent
papers,~\cite{luki}, Lukierski \ETAL  developed a theory of the deformation
of the Poincar\'{e} group. In order to accomplish the task, the
authors of ref.~\cite{luki} started from the
DeSitter group. They performed a group deformation along the lines
described in the classic papers on the deformation of simple groups,~
\hbox{\cite{drin}, \cite{man},\cite{jimbo}} and afterwards they performed
a  contraction
to a deformed Poincar\'{e} group. While the procedure referred to
is not a unique one, (see, \EG ~\cite{oleg}) it is internally consistent.
Moreover, it is attractive from the physical point of view:
the theory contains a characteristic mass, $\kappa$. The latter
determines the energies below which
the special theory of relativity is valid to a good approximation;
at energies above $\kappa$, significant deviations from the familiar
relativistic kinematics are to be expected. The important point is
that on can obtain useful information about the possible deformation
of the Poincar\'{e} group from the kinematics of a single particle alone.
(This is due to the fact that the translation subgroup remains an
Abelian one: the rules of the addition of momenta are unchanged.)

Existing and future observational data on ultra high energy (UHE),
distant events can place an interesting lower limit on the characteristic
mass, $\kappa$. For this purpose one needs, in essence,
only one of the results
developed in refs.~(\cite{luki}).  In those papers it was shown
that the quadratic Casimir invariant of the Poincar\'{e} group
is deformed to the following expression:
\BEQ
C_{\kappa} = - \left( {\bf p}^{2} + 2 \kappa^{2} \left( \cos
\frac{E}{\kappa}
- 1 \right) \right).
\label{eq:cas}
\EEQ
In  \eq{eq:cas} the quantities ${\bf p}$ and $E$
have their usual meaning.
The deformed Casimir invariant, \eq{eq:cas}
goes over to the familiar quadratic form
as $\kappa \rightarrow \infty$. However, at any
$\kappa < \infty$, there is an observable difference between
ordinary relativistic kinematics and the one described by
\eq{eq:cas}. Let us consider, in particular, a single
particle state, defined, as usual, as an irreducible
representation of the deformed Poincar\'{e} group.
In that case, $C_{\kappa}=m^{2}$, where $m$ is related
to the the rest energy of the particle. On putting ${\bf p} = 0$,
the rest energy is given by:
\BEQ
E^{0} = \kappa \cos^{-1}\left( 1 - \frac{m^{2}}{2 +
\kappa ^{2}}\right),
\EEQ
which agrees with the usual expression up to corrections of
$O\left( m^{4}/\kappa ^{4}\right)$.
One can compute the group velocity ($v$) of the
particle in the deformed Poincar\'{e}
group. As whithin the framework of the usual (undeformed) Poincar\'{e}
group, if the energy of the particle is much greater than $m$,
the latter can be neglected.
In this  approximation one has:
\BEQ
\frac{1}{v} = \frac{d\left| {\bf p}\right|}{dE}
\approx \cos\frac{E}{2\kappa}.
\label{eq:groupv}
\EEQ

The deformed Poincar\'{e} group leads to the result that
a massless particle is, in the usual sense of the word, always
{\em superluminal}: what one accepts these days as the ``speed of light''
is only the {\em low frequency limit} of the
group velocity.

One can now place a lower limit on the value of $\kappa$ as follows.
Suppose that a periodic UHE signal emitted by a
distant source, (\EG  a pulsar) is observed.

The time delay a $\delta$ function pulse suffers between a source
a distance $d$ away from a (terrestrial) detector is given by the
elementary formula:
\BEQ
\Delta t = \frac{d}{v} = d \cos \left( \frac{E}{2 \kappa}\right)
\label{eq:delta}
\EEQ

We represent the signal by means of a Fourier series. For a
signal of fundamental period $T$ and energy $E$, one has at the source:
\BEQ
I(E,t_{s}) = \sum_{n= -\infty}^{\infty} I_{n}^{S}(E) \exp i n \omega t_{s},
\label{eq:source}
\EEQ
where  $\omega = 2\pi /T$.
The intensity is real, $ I_{-n} = I_{n}^{*}$.
At the detector, the Fourier coefficients differ from the ones at
the source by an energy dependent phase shift, analogously to the
time delay in \eq{eq:delta}:
\BEQ
I_{n}^{D}(E) = I_{n}^{S}(E)
\exp\left[i \omega n d \cos \left( \frac{E}{2\kappa}\right) \right]
\label{eq:detector}
\EEQ

We now average \eq{eq:detector} over the energies, assuming, as
usual that the spectrum
of the source is given by a power law:
\BEQ
\frac{d \Phi}{d E}= C E^{-\alpha },
\EEQ
where $C$ is a normalization constant in order to assure
$\int_{E_{min}}^{E_{max}}\left( d\Phi/dE\right)dE =1$.
Here, the quantities $E_{min}$ and
$E_{max}$ stand for  the minimal and maximal energies observed.
The averaging over energies is made necessary by the fact that one has no
direct information on the pulse shape at the source.
We have:
\BEQ
\langle I_n \rangle = C \int_{E_{min}}^{E_{max}} I_n^S\left( E\right)
E^{-\alpha}\exp\left[i \omega n d \cos \left( \frac{E}{2\kappa}\right) \right]
\label{eq:avg}
\EEQ

As a consequence of the velocity dispersion,
both the magnitude and the
phase of $\langle I_{n}^{D}\rangle$ differ from $I_{n}^{S}$.
 On assuming that the pulse at the source
is proportional to a $\delta$ function at all energies
--- \IE  that the entire observed width of the  pulse arises
from velocity dispersion --- one can obtain a {\em plausible lower limit}
on $\kappa$. (Evidently, this assumption means that $I_{n}^{S}$ is
independent of $n$ and $E$.)
The argument goes as follows.

Assuming that the pulse is a $\delta$ function
at the source at any energy, one shows in a straight--forward manner that
the signal observed at the detector has a finite width as a consequence
of the assumed velocity dispersion: one just has to insert
$I_n^{S}= I = \mr{const}$  into \eq{eq:avg}. The reason why such an
argument cannot be formulated as  a rigorous theorem is that one can
construct counterexamples showing that a signal emitted
at a source can actually become narrower at arrival. (For instance,
one can imagine that the high energy particles are emitted after
the low energy ones and
 just catch up with them at the detector.
Thus the pulse observed at the detector is narrower than the one
at the source.
Examples of this kind have been analyzed in detail by \cite{cudell}.)
One has to note, nevertheless that all scenarios of the type mentioned above
are highly ``unnatural'': they require a delicate
``fine tuning'' of the pulse at the source according to its distance
from a detector at Earth. (In particular, there is no known acceleration
mechanism which would produce effects similar to the one described above.)
For this reason, we do not consider mechanisms of this type any further.
In order to complete the argument, we note that the pulse has a finite natural
width at the source. Hence, any broadening caused by a velocity
dispersion is superposed onto the natural width. Consequently, even in the
total absence of a velocity dispersion, one would see a finite width and,
hence, estimate that $\kappa < \infty $.

The unknown coefficient $I$ is eliminated by taking ratios of the moduli of
the Fourier coefficients. It is evident from these equations that in order to
obtain good lower limits on $\kappa$,
\begin{itemize}
\item one has to maximize the distance between the source and the detector,
\item one has to use data at the highest energies available.
\end{itemize}

There was a burst observed in the binary system HER--X1 in 1986. The
distance of this system is about 5 kpc;  particles of the highest
energies were observed by the CYGNUS collaboration by means of
detecting  extensive air showers induced by them,
\CF  ref.~\cite{cygnus}. The nominal threshold of the detector
as stated in ref.~\cite{cygnus} is 70 TeV. Using the period--folded data
from this experiment and the published fundamental period,
($T\approx 1.24$ sec), ref.~\cite{folded}, we determined the first few
Fourier coefficients from the data,~\cite{Fourier}.

(Due to the fact that 11 showers
were observed by the CYGNUS collaboration, the data allow for a meaningful
determination of the first five complex Fourier coefficients. The
coefficient $I_{0}$, corresponding to a constant intensity, can always be
subtracted.) The result is displayed in Table 1.
\vskip5mm
\[
\begin{array}{|c|c|c|}
\multicolumn{3}{c}{Table \quad 1: \mr{Fourier coefficients}} \\
\hline\hline
n & \left|I_{n}^{D}\right|^{2} & \arg I_{n}^{D} \\
\hline
1 & 0.99 & 0.044 \\
2 & 0.96 & 0.088 \\
3 & 0.92 & 0.13 \\
4 & 0.86 & 0.17 \\
5 & 0.79 & 0.21 \\
\hline
\end{array}
\]
\vskip5mm

One observes that the phases of the Fourier coefficients determined from
the data
exhibit no noticeable trend; probably, they are considerably
affected by ``noise'', \IE   by the various errors in the data set.
This is a general phenomenon~\cite{alex}. For this reason,
we used only the moduli of the Fourier coefficients in order to obtain a
lower limit on $\kappa$. (One determines by inspection that the moduli
contain less noise than the phases: they are decreasing almost
linearly with $n$.)

Using the ratio, $\left| \langle I_{2}^{D}\rangle /\langle I_{1}^{D}
 \rangle
\right|^{2}$ from \eq{eq:avg}, one can compute  $\kappa$.
As stated before, this is a lower limit on its value. The calculation
cannot be carried out in closed form: for values of $\kappa, E_{min}
\mr{and} E_{max}$ of interest, the integral cannot be well approximated
by any closed expression in terms of known functions. On taking
the minimal and maximal energies from ref.\cite{cygnus}, the integral
was computed numerically for various values of $\kappa$. Then, taking the
the first four ratios of the moduli of the observed Fourier
coefficients, the ``optimal'' value of $\kappa$ was determined
by minimizing $\chi^2$ between the observed and computed values.

The function $\chi^{2}(\kappa)$ has a rather broad minimum at
\BEQ
\kappa  \approx 1.3 \times 10^{12} \mr{GeV}.
\label{eq:lowerlimit}
\EEQ
Due to the fact that
the determination of the primary energy of a shower is uncertain to
about a factor of 2 (cf. \cite{cygnus}), it is of interest to
test the sensitivity of the value (more precisely, the lower limit)
of $\kappa$ to the primary energy.  Inspection of \eq{eq:avg}
shows that the the integral is more sensitive to variations
of $E_{max}$ than to $E_{min}$. Indeed, varying $E_{min}$ by a factor
of 5, we found no sensitivity (to within 3\%) of the lower
limit of $\kappa$. In Fig.(1), we plotted
$\chi^{2}$ as a function of $E_{max}$ and $\kappa$. It is evident that
neither the lower limit of $\kappa$, nor the value of $\chi^2$ at
minimum depend sensitively on the primary energy of the most energetic
shower.
In arriving at the result \eq{eq:lowerlimit}, we used a value,
$\alpha = 1.4$, obtained from a rough fit to the data. The quoted lower
limit is rather insensitive to the precise value of the spectral index.

Previous to our work, the authors of ref.~\cite{folded}  quoted
a lower limit on the characteristic mass scale(s) of the {\em
ad hoc} modifications of relativistic kinematics,~\cite{SN1987A}.
They find a result which is equivalent in the present
formulation
to $\kappa \gappr 5  10^{11}$~GeV\footnote{In order to compare the estimate
used in ref.~\cite{folded} with ours, one has to approximate
$\cos E/\kappa$ in \eq{eq:delta} by the first two terms of its
Taylor series.}. The improvement
represented by \eq{eq:lowerlimit} is due to the fact that by
computing Fourier coefficients, one uses a larger amount of
information contained in the data than by using the simpler
estimates in ref.~\cite{folded}. In fact, one sees from
Fig.(1) that the value of $\kappa $ quoted in
ref.~\cite{folded} gives a $\chi^2$ about  3 times as high as
its minimal value.
More importantly, the
the formula used in ref.~\cite{folded}, depends quadratically
on $E_{max}$ and, since $E_{min}\ll E_{max}$, it is practically
independent of the least energetic shower. Thus, in effect,
the lower limit is determined by a single event only. By contrast,
in the method described here, not  only the function $\chi^{2}(\kappa)$,
but individual Fourier coefficients as well, show a much weaker
dependence on the energy of an individual shower. Consequently,
even if one or a few of the showers came from the background,
the estimate of the lower limit will not be changed significantly.

It appears that deviations from Poincar\'{e}
invariance are negligible at least up to energies which
are fairly close to a presumed grand unification  mass.
Presumably, as larger data samples
become available, this limit can be further improved, or else,
 one may be able to estimate the energy where the
special theory of relativity breaks down due to some new physics.
 UHE astrophysical data are unique in
this respect due to the high energies and large distances involved.
\vskip6mm
\ack .\hskip1em  We also thank
Alex Szalay for useful discussions.
\begin{sloppypar}

\end{sloppypar}
\newpage
\CLI{\large\bf Figure Caption}
\vskip5mm
Fig.(1) The quantity $\chi^2$ is plotted as a
function of $\kappa$ end of the maximal observed shower energy.
$E_{max}$ is plotted in PeV; $\kappa$ is plotted in units of
$5\times 10^{11}$GeV.
\end{document}